# An Omnidirectional Approach to Touch-based Continuous Authentication

Peter Aaby[1], Mario Valerio Giuffrida[1], William J Buchanan[1], Zhiyuan Tan[1]*
[1]Edinburgh Napier University, School of Computing, Edinburgh, United Kingdom
*Corresponding author: Zhiyuan Tan
E-mail: z.tan@napier.ac.uk

**Abstract**

This paper focuses on how touch interactions on smartphones can provide a continuous user authentication service through behaviour captured by a touchscreen. While efforts are made to advance touch-based behavioural authentication, researchers often focus on gathering data, tuning classifiers, and enhancing performance by evaluating touch interactions in a sequence rather than independently. However, such systems only work by providing data representing distinct behavioural traits. The typical approach separates behaviour into touch directions and creates multiple user profiles. This work presents an omnidirectional approach which outperforms the traditional method independent of the touch direction - depending on optimal behavioural features and a balanced training set. Thus, we evaluate five behavioural feature sets using the conventional approach against our direction-agnostic method while testing several classifiers, including an Extra-Tree and Gradient Boosting Classifier, which is often overlooked. Results show that in comparison with the traditional, an Extra-Trees classifier and the proposed approach are superior when combining strokes. However, the performance depends on the applied feature set. We find that the TouchAlytics feature set outperforms others when using our approach when combining three or more strokes. Finally, we highlight the importance of reporting the mean area under the curve and equal error rate for single-stroke performance and varying the sequence of strokes separately.

*Keywords:* Behavioural Biometric; Continuous Authentication; Touch Biometric; Smartphone Security; Model Selection

## 1. Introduction

In 2007, Apple caused a paradigm shift by releasing its first smartphone with a touch screen. Since then, smartphones have become ubiquitous, with an 81% penetration rate in the US [1]. With the adoption of smartphones, a single device can now provide access to the entire life of its owner, e.g., entertainment profiles such as Netflix, social media accounts with instant messaging, and online banking, amongst others. However, user authentication on touch devices is challenging due to the limited input interfaces. With facial recognition and smartphone fingerprint reading, biometric lock screen authentication can confirm legitimate users conveniently but cannot continuously maintain user authenticity through user sessions. These physiological biometrics also require sensors vulnerable to presentation and replay attacks [2], [3]. Finally, active authentication methods are time-consuming and may interrupt or delay productivity [4], [5].

Continuous Authentication (CA) mitigates these weaknesses by passively collecting behavioural biometrics from user input and evaluating user authenticity passively over time. Figure 1 visualises the concept of touch-based CA. Step 1 captures touch data, Step 2 compares against known touch behaviour, and Step 3 evaluates and produces a lock or unlock decision, Steps 3.1 and 3.2, respectively. The continuous loop then repeats the steps for each interaction to secure the device over time, continuously

The early work by [6] sought to establish the viability of touchscreen data as input for behavioural biometrics and explicitly argued for user authentication through touch strokes. Contrary to other types of CA, the touch-based method only requires a touchscreen and may thus be applied across any device with a touch interface. For instance, humans in smart factories could use a touchscreen to operate a conveyor belt or pickers in a warehouse to use a smartphone for packing orders. However, several challenges remain, such as identifying high-quality behavioural traits and defining a standard to compare touch-based CA methods [7], [8].

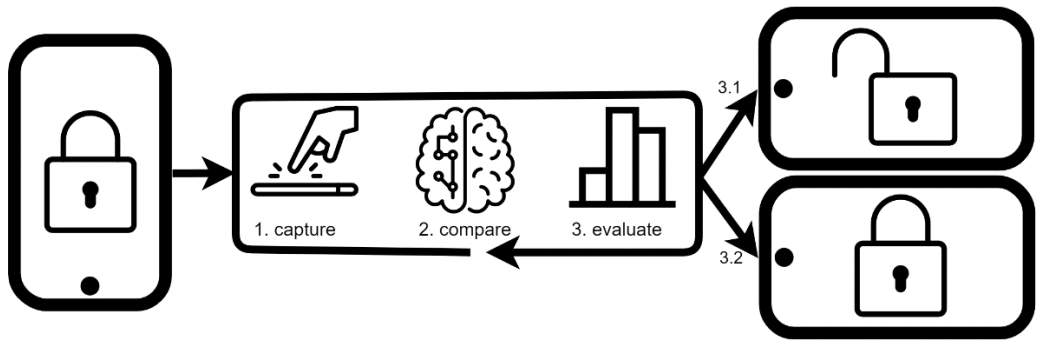

*Figure 1 Touch-based continuous authentication concept. In this example, our implementation secures a phone by 1. capturing biometric touch data; 2. comparing against a biometric user template; and 3. evaluating whether the genuine user is granted access to the phone 3.1 or locked 3.2*

### 1.1. Motivation and challenges

Commonly, researchers model touch behaviour in the context of Horizontal strokes (Hs) or Vertical strokes (Vs), with the Hs model typically outperforming Vs [6], [9], [10]. Figure 2 presents the typical modelling approach for a user touching their device four times. Each stroke is first evaluated for direction and then provided to the respective bidirectional model. The model then predicts a probability score between zero and one for strokes 1 through 4. Higher probability results define a favourable decision to unlock the device - as there is a high probability that the user is genuine.

As further explained in sections 2.3 and 3.8, a moving average can be applied as a window over the strokes and probabilities to smoothen and improve authentication accuracy. However, we suggest that a model may be agnostic to the direction; thus, a single *omnidirectional* model should authenticate the owner regardless of touch direction. A recent study by [10] took early steps towards investigating distinct traits concerning individual user behaviour. However, their work focused on limited features and did not consider mixing the directional gestures into a single omnidirectional model. Consequently, this paper is motivated by the challenges and discrepancies between modelling approaches, the lacking comparison of behavioural feature sets, and a desire to create user profiles based on high-quality behavioural features. Lastly, we define a new method to select the model parameters to reduce complexity at the cost of minor performance drops.

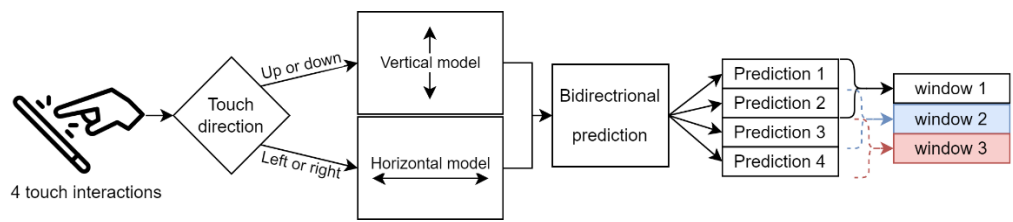

*Figure 2 A high-level overview of the traditional bidirectional approach. Commonly, interactions are broken into touch-direction with individual models to predict if a user is genuine. Further improvements are made to the performance by combining predictions using sliding windows, among others.*

The following section will detail each contribution related to these challenges.

### 1.2. Research questions and contributions

Following the challenges outlined, this paper seeks to contribute answers to the following questions:

1. What is the performance difference between the proposed approach versus a typical bidirectional model where stroke orientation is separated and parameters are highly optimised?
2. What is the impact of combining *n*-strokes when using the typical and proposed approach?
3. Which feature set should be used considering the different directional modelling approaches?

We focus on balancing the model complexity and performance to answer these questions while choosing the optimal feature set for a single omnidirectional model and two independent Hs and Vs models. Each user is modelled using five feature sets to evaluate the best over-all behavioural traits. We report the Area Under the Curve (AUC) since it is threshold independent while assessing the best trade-off between classes as a function of all thresholds [11]. We also report the Equal Error Rate (EER) since it is the most popular metric across the literature, noting that such a rate only represents a specific decision threshold. Finally, our results are reported for single-stroke and combining strokes, measured by the AUC and EER scores when combining a sequence of strokes in ranges 1 through 20.

The rest of the paper is structured as follows: Section 2 covers the related work. Section 3 describes the experimental design and the applied methods to implement and complete the experiment. Section 4 presents the results before concluding in Section 5.

## 2. Related work

Touch-based continuous authentication relies on distinct features to authenticate an owner from other users. Several approaches from the literature have exhibited promising results using different feature sets, although some overlap where others use existing feature sets from the literature [6], [9], [12]–[14]. This work focuses on five different feature sets chosen because they offer the best variation amongst the related literature. The study presented in [12] explains how to differentiate between child and adult smartphone users using touch-based features. While their feature set is not used in authentication, the features they extract could also define distinct behavioural features used for authentication; thus, applying these in continuous authentication would be attractive. The authors conclude that strokes perform better than clicks in differentiating children from adults and that an Extra Tree (ET) classifier outperforms others. The ET classifier is not commonly used for CA and could also be interesting in the context of CA. The following sub-section will further compare the classifiers often used for CA.

### 2.1. Classifier, parameters, and metrics

Rather than identifying children from adults, Table 1 presents work that uses various classifiers to implement touch-based CA and specifies the varying parameters used across the literature. Although each classifier behaves differently, they can also differ internally depending on parameter settings. The CA literature often evaluates several classifiers with varying parameters [9]. However, it is challenging to characterise the best overall classifier amongst the related literature without specifying comparable parameter search space or using the same metrics. For example, [6] achieves a 13% EER on a single stroke using a Support Vector Machine (SVM) with a Radial Basis Function kernel. However, [15] finds Random Forest (RF) superior, with a single stroke accuracy of 65%. In [16], the RF classifier offers the best EER score of 25% instead. Finally, [13] optimises for a balanced F-score with a single stroke performance between 0.7 and 0.8 - depending on the type of gesture. Thus, the papers are challenging to compare beyond their different approach as different metrics are often used.

Table 1 Classifiers and parameters across related work: k-Nearest Neighbors (KNN), Logistic Regression (LR), Naïve Bayes (NB), Multi-Layer Perceptron's (MLP), Bayesian Network (BN), Decision Tree (J48), Back Propagating Neural Network (BPNN), One-Class SVM (OCSVM), Isolation Forest (iForest). Undefined parameters are marked with N/A.

| Paper | Classifier (parameter) |
|-------|------------------------|
| [6]   | SVM ($C$=N/A, $\gamma$=N/A), KNN ($k$=1,3,5,9) |
| [9]   | SVM ($C$=N/A, $\gamma$=N/A), KNN ($k$=9), RF ($n$=1000), LR, NB, MLP, J48 |
| [12]  | SVM ($C$=1.4, $\gamma$=0.15), KNN ($k$=7), RF ($n$=200), ET ($n$=200) |
| [15]  | SVM ($C$=2,8, $\gamma$=8), KNN ($k$=3), RF ($n$=10,100) |
| [16]  | SVM ($C$=0.03, $\gamma$=0.006), KNN ($k$=2-20. Best=11), RF ($n$=1000), BPNN (2+1 layers and *learning rate*=0.001) |
| [13]  | OCSVM ($nu$=0.1), iForest (*contamination*=0.1) |

### 2.2. Modelling approach and the impact of training size

When modelling touch-based CA, strokes can be categorised and processed depending on the direction of the trajectory or independent of the direction. For example, grouping left, and right strokes in a Hs model are used in [6], [9], [13], [15], [16]. In contrast, [16] models each direction individually while evaluating against mixed directions. These papers approach the classification as a binary challenge using a One versus Rest (OvR) scheme [17]. The device owner then forms the positive class, and the negative class groups the remaining users. OvR causes a class imbalance that can be mitigated through sampling techniques [6], [9]. However, it becomes increasingly challenging to compare works since the directional approach differs, and OvR sampling may further affect the characteristics of training data. Thus, we highlight the varying amount of training data used and the potential effect on performance. Eighty samples are used for Hs/Vs models in [9], 100 for Hs/Vs in [15], and roughly 160 per direction-specific models in [16], for each class, respectively. The concept of model stability through varying training data size is partially studied in [14], [15], with [13] showing minor improvement using more than 80 training observations.

### 2.3. Removing clicks and combining strokes

Defining strokes from clicks is essential as a precursor to modelling since clicks appear to cause poor performance [13]. There are different ways to identify strokes, e.g., counting the points within a trajectory and removing strokes with less than four [9] or five [6] touchpoints. Others assess the directional angle and exclude strokes that change direction, such as sliding up and then down without releasing the finger [13], [14], [16], or a minimum length can be required [15]. Since a user's touch strokes may have slight variations, authenticating based on a single stroke is challenging because it requires the classifier to identify each touch operation perfectly. In [6], a range of 1-20 strokes are combined using different techniques based on KNN neighbours' distance and the SVM's hyperplane, with 11 and 13 strokes working well. In contrast, [9] takes a sequence of ten feature vectors and applies a moving average when predicting the user. Rather than averaging the feature vector, [15] uses a moving average over the predicted probabilities of a sequence of strokes and concludes that ten strokes are optimal. The latter approach can be seen in Figure 2 with a moving average window of two strokes. However, [16] found 11 strokes a reasonable trade-off. [14] groups strokes by five and authenticated based on a majority vote. Lastly, [12] combines 9-11 strokes. Consequently, improving performance by combining around ten strokes is common, but the method and outcome vary across the literature. As such, we seek to answer research question two by varying the number of combined strokes in the context of the different approaches and feature sets.

The following section presents the experimental design and describes the proposed omnidirectional method, training size, data set, and the implemented behavioural feature sets.

## 3. Experimental design and methods

The central hypothesis of this paper argues that behaviour can be generalised by an omnidirectional model - matching or outperforming the traditional approach where the horizontal and vertical strokes are modelled independently. If true, the time to model a user can be reduced by roughly half. Further, selecting and evaluating essential behavioural features may be more straightforward as only one model needs to be inspected. The traditional approach is configured as a baseline and omnidirectional as the contender to evaluate our method. We define an omnidirectional model to process any gesture independent of the direction of the stroke. In contrast, bidirectional models separate strokes depending on the underlying direction. Furthermore, five different feature sets are used to illuminate which behaviour works in the context of the proposed method.

### 3.1. Data and feature sets

This paper uses the raw data collected by [9] as it contains more users, observations, and extended periods of data compared to others [6], [15], [18]–[22]. The data was collected over two sessions to enable intersession authentication. The raw data includes a user ID, swipe ID, timestamp, ($x, y$) coordinate pairs, pressure, and the area covered by the finger for each recorded touch point. Portrait and landscape data are separated into different sets. We exclusively focus on the portrait mode as the most preferred smartphone orientation [23]. In Figure 3, a single stroke is visualised and shows a user moving their finger from Point A, the right side of the screen and leftwards to Point B.

### 3.2. Compatible users and raw data

Since this work is not looking to vary the training data, we follow the recommendation by [13]. Thus, eligible users must provide more than 80 training observations with each horizontal and vertical direction to establish the two models for the bidirectional approach. An even number of observations is selected for each direction among up, down, left, and right strokes. In contrast, the omnidirectional model uses all directions. When authenticating, users must also provide enough test data to combine a sequence of $n$ strokes. [6] combined up to 20 strokes with ten strokes producing good results; thus, we combine strokes in the range of 1 through 20 for comparability.

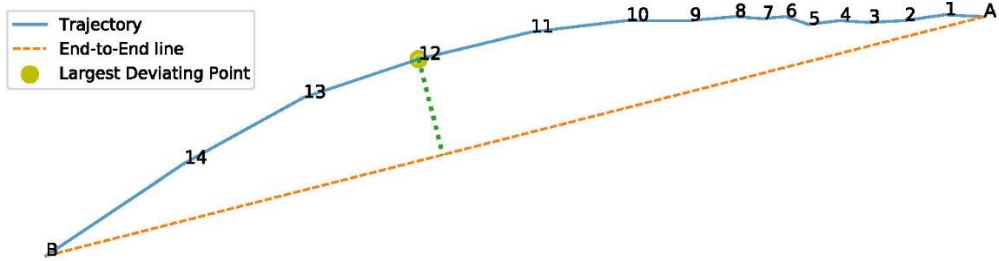

*Figure 3 Example of raw data from a single stroke. Each touch point describes behavioural traits through engineering features such as "length of trajectory", "End-to-End line", and the "Largest Deviating Point" between the two feature examples.*

### 3.3. Features

Figure 3 presents a single stroke drawn between a finger down and up, Point A and Point B, respectively. A blue line is illustrated as the trajectory, which defines the length between the 16 points collected as part of the Android operating system when capturing touch interactions. An orange dotted line is also shown to highlight the End-to-End length (E2E) feature, which defines the flight distance. The Largest Deviating Point (LDP) is another feature that appears at point 12, measured by the dotted green line. Several feature sets have been proposed throughout the literature to describe behaviour, as seen in Table 2. Interestingly, several papers overlap without direct comparison or discussion on which features are included or excluded in each set.

### 3.4. Feature extraction and data cleaning

We selected five papers [6], [9], [12]–[14] from the literature because they provide a broad spectrum of different behavioural traits. To focus on strokes, we remove clicks and interactions with less than or equal to five points or if the trajectory length is shorter than three pixels. Besides filtering clicks, some features cause undefined values, such as the E2E line slope for perfect horizontal stroke. Similarly, the inter-stroke time is unavailable for each user's first stroke. After removing strokes and data cleaning, the data set consists of 78,423 strokes that qualify for all five feature sets.

*Table 2 Overview and overlap of the feature as shown in Frank et al. [6], Serwadda et al. [9], Syed et al. [14], Yang et al. [13], and Cheng et al. [12]. Most papers have several features overlapping with the earliest work by Frank et al., whereas Yang et al. have only four overlapping features and limited overlap with others.*

| # | Feature name | [6] | [9] | [14] | [13] | [12] | # | Feature name | [6] | [9] | [14] | [13] | [12] |
|---|---|---|---|---|---|---|---|---|---|---|---|---|---|
| 1 | Inter-stroke time | x | | x | | x | 39 | Standard deviation pressure | x | | | | x |
| 2 | Stroke duration | x | x | x | | x | 40 | 25% pressure | x | | | | |
| 3 | Start X | x | x | x | x | x | 41 | 50% pressure | x | | | | |
| 4 | Start Y | x | x | x | x | x | 42 | 75% pressure | x | | | | |
| 5 | Stop X | x | x | x | x | x | 43 | Mean area | x | | | x | |
| 6 | Stop Y | x | x | x | x | x | 44 | Standard deviation area | x | | | | x |
| 7 | Length E2E | x | x | x | | x | 45 | 25% area | x | | | | |
| 8 | Mean resultant length | x | | | | x | 46 | 50% area | x | | | | |
| 9 | Numeric direction | x | | | | | 47 | 75% area | x | | | | |
| 10 | Direction E2E | x | x | x | | x | 48 | Start pressure | | | x | x | x |
| 11 | 20% velocity | x | | x | | | 49 | Stop pressure | | | x | x | |
| 12 | 50% velocity | x | x | x | | | 50 | Categorical direction | | | x | | |

| # | Feature name | Paper [6] | [9] | [14] | [13] | [12] | # | Feature name | Paper [6] | [9] | [14] | [13] | [12] |
|---|---|---|---|---|---|---|---|---|---|---|---|---|---|
| 13 | 80% velocity | x | | x | | | 51 | X @ max velocity | | | | x | |
| 14 | 20% acceleration | x | | | | | 52 | X @ min velocity | | | | x | |
| 15 | 50% acceleration | x | x | | | | 53 | Y @ max velocity | | | | x | |
| 16 | 80% acceleration | x | | | | | 54 | Y @ min velocity | | | | x | |
| 17 | Median velocity last 3 pts | x | | | | x | 55 | Max velocity | | | | x | x |
| 18 | Largest deviation from E2E | x | | | | | 56 | Min velocity | | | | x | |
| 19 | 20% deviation | x | | | | | 57 | Slope of E2E line | | | | x | |
| 20 | 50% deviation | x | | | | | 58 | Intercept of E2E line | | | | x | |
| 21 | 80% deviation | x | | | | | 59 | X @ LDP | | | | x | |
| 22 | Average direction | x | | | | | 60 | Y @ LDP | | | | x | |
| 23 | Length of trajectory | x | x | x | | x | 61 | LDP pressure | | | | x | |
| 24 | Ratio length E2E-to-trajectory | x | | x | | | 62 | Mean velocity X-axis prev to LDP | | | | x | |
| 25 | Mean velocity | x | x | x | | x | 63 | Mean velocity Y-axis prev to LDP | | | | x | |
| 26 | Median acceleration last 5 pts | x | | | | | 64 | Mean velocity X-axis post to LDP | | | | x | |
| 27 | Mid-stroke pressure | x | | x | | x | 65 | Mean velocity Y-axis post to LDP | | | | x | |
| 28 | Mid-stroke area | x | | | | x | 66 | Start pressure | | | | | x |
| 29 | Mid-stroke finger orientation | x | | | | | 67 | Time to reach max velocity | | | | | x |
| 30 | Phone orientation (label) | x | | | | | 68 | X displacement finger down-down | | | | | x |
| 31 | Standard deviation velocity | | x | | | x | 69 | Y displacement finger down-down | | | | | x |
| 32 | 25% velocity | | x | | | | 70 | X displacement finger down-up | | | | | x |
| 33 | 75% velocity | | x | | | | 71 | Y displacement finger down-up | | | | | x |
| 34 | Mean acceleration | | x | | | | 72 | Median velocity first 3 pts | | | | | x |
| 35 | Standard deviation acceleration | | x | | | | 73 | Mid-stroke velocity | | | | | x |
| 36 | 25% acceleration | | x | | | | 74 | Median acceleration first 3 pts | | | | | x |
| 37 | 75% acceleration | | x | | | | 75 | Median acceleration last 3 pts | | | | | x |
| 38 | Mean pressure | | x | | | x | 76 | Mid-stroke acceleration | | | | | x |

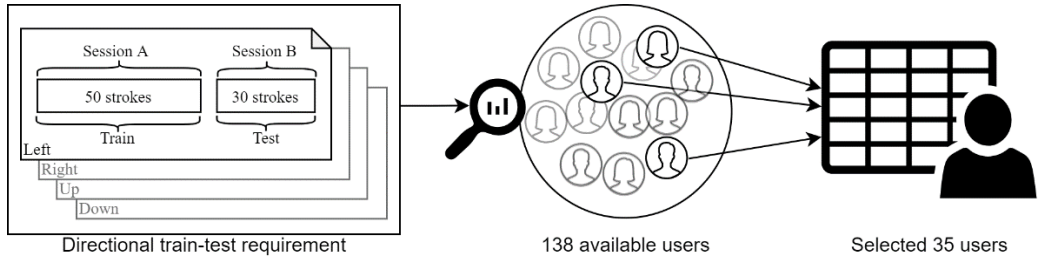

*Figure 4 Subsetting data and selecting users with adequate training and testing data. Each user is balanced to include a fixed number of directional training and testing interactions to avoid bias towards a specific touch direction. The overall data set is reduced from 138 candidates to 35 valid users with sufficient data.*

### 3.5. Selecting users of interest

Each model must have 80 observations to generate a stable behavioural model for the target user [13]. Thus, valid users are chosen based on the requirement seen in Figure 4. For each left, right, up, and down direction, 50 training and 30 testing strokes are required. The training size is thus 100 for each of the two bidirectional models and 200 for the omnidirectional. The testing size is set to 30 samples per direction to ensure enough data when combining strokes. Consequently, the data set is reduced from 138 to 35 users of interest.

### 3.6. Modelling pipeline and class balancing

In Figure 5, the proposed modelling pipeline is introduced, where data is labelled according to the OvR method for each user. A Cross-Validator (CV) is configured using a five-fold stratified loop, repeated five times, and is applied to reduce bias considering the limited number of training observations. When training the bidirectional model using 100 samples and the CV, each training fold is reduced to 80 strokes which adheres to the guidelines of [13]. The pipeline is implemented using Python 3.8 SciKit-learn [24] and ImbalancedLearn [25] as the sampler. The touch training data from Session A is provided to the pipeline for each user and the specific classifier per feature set from Table 2. Figure 5, Step 1, undersamples the majority class. Steps 2 and 3 standardise to zero mean and normalise values between 0-1 if the classifier requires it, e.g., SVM. Finally, Step 4 implements the parameters in Table 3.

Once a model is computed, the testing data from Session B is evaluated by the model, generating predicted probabilities. Data from Session B is collected at least one day after Session A. Thus, the results measure the intersession performance and are possibly more conservative than works measuring intrasession performance [6], [9], [10]. Because of the CV, the pipeline defined in Figure 5 computes 25 (= 5x5) models and searches for the optimal parameter for each feature set, classifier, and approach. Thus, the parameter grid in Table 3 results in a broad range of models to evaluate, such as an ET classifier, and the optimal parameter is selected by considering a total of 236,250 models. Since class imbalance can be challenging for some classifiers, the proposed method down-samples the majority class following an OvR approach similar to [9], [14]. At the same time, stratification ensures class balance in each fold.

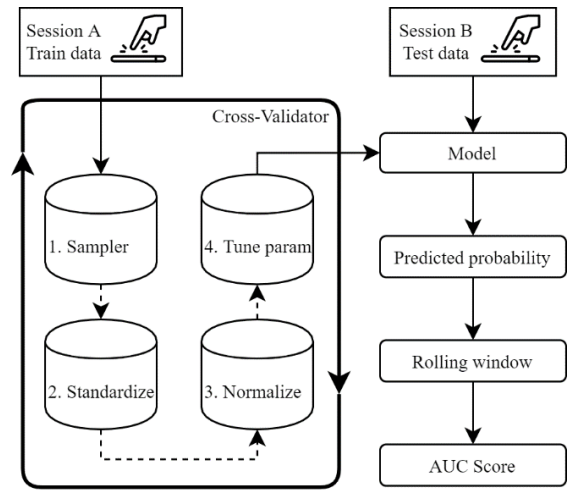

*Figure 5 Modelling pipeline used for training. Each user provides their touch data from session A for training, which is then 1. downsampled for OvR class balance, 2. scaled to one standard deviation and zero mean, 3. normalised in the range 0-1, and 4. tuned against our parameter list in Table 3. Step two and three is not required for the tree-based classifiers. The cross-validator thus creates models for each user that can predict user input from session B for each algorithm and feature set. An average rolling window is applied to the predicted probabilities to smoothen and increase general performance.*

### 3.7. Classifier parameters and complexity

This work assesses whether the proposed single omnidirectional model can compare with bidirectional models using different feature sets and classifiers. However, each classifier may perform differently depending on the configured parameters and the provided feature set. Classifiers such as KNN, SVM, RF, and ET are commonly seen in the literature with different parameters [6], [9], [12]–[14]. Still, in published work, it can be unclear which parameters are tested and thus the most effective across the different behavioural feature sets. To address this, we implement each classifier while searching the parameters shown in Table 3.

*Table 3 Parameter search space for each classifier tested, as seen in the literature. The parameters are chosen based on the related work where applicable.*

| Classifier | Parameters |
| --- | --- |
| KNN | K = {1, 3, 5, 7, 9} |
| SVM | C = {0.01, 0.1, 1.0, 10, 20, 100} |
| GB / RF / ET | Min samples split = {0.005, 0.01, 0.1} |
| | N estimators = {100, 200, 500, 700, 1000, 1200} |

Most classifiers become more complex as their parameters increase in value, such as the number of trees in RFs and the regularisation parameter of SVM- which may lead to overfitting models. Thus, a consideration is made between balancing the best parameter and the model complexity. The parameters are optimised to maximise the Area Under the Curve (AUC) score as a function of all thresholds [11]. We select the classifier's parameter by subtracting one standard deviation of the AUC score from the model's best-performing AUC score, trading minimal performance gains for reduced complexity.

Figure 6 visualises our parameter selection approach. The example starts with the cross-validated output of three parameters tested for a given algorithm. From the three test results, Rank 1 provides the highest AUC score. However, the standard deviation value is often high while providing minor performance over the other results. Thus, we take the best AUC score and subtract the associated standard deviation value to set a threshold of the test results, which defines a mask of acceptable parameters. This example has two parameter pairs as the mask, in which the lowest parameter is selected since it produces a less complex model while generally preserving good performance. Consequently, we sacrifice minor performance while lowering the deviation between users. Similarly, it reduces the model complexity, translating to faster training.

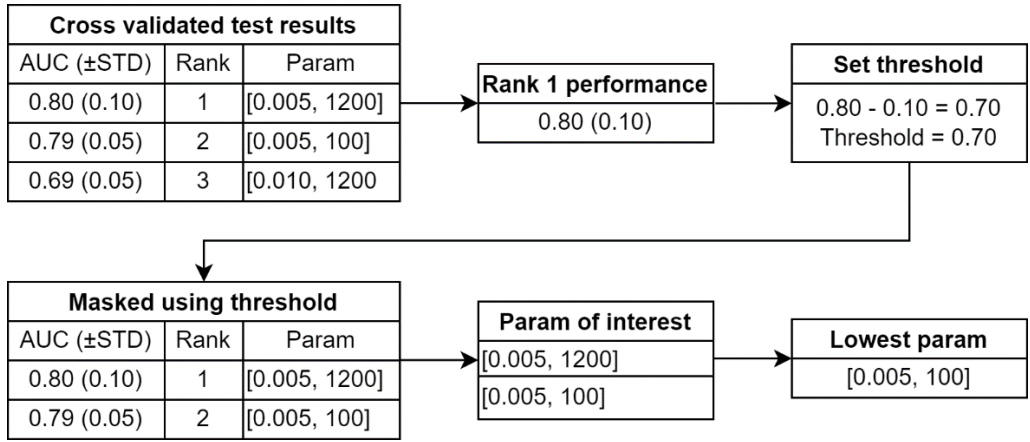

*Figure 6 An example of the parameter selection method when balancing performance and complexity. Step 1. locates the highest-ranked AUC score; 2. computes a lower threshold by subtracting the associated STD; 3. uses the threshold to mask the cross-validated parameters of interest; and 4. then selects the lowest parameters, which create lesser complex models.*

While classifiers may have additional parameters that can affect performance, the scope of this paper is to evaluate those tuned in the related work for comparability. The details and mechanisms of each classifier are well documented across the literature, but the following parameters are briefly covered. For SVM, the kernel used is a Radial Basis Function and the $\gamma$ parameter scales according to the number of features and their variance [24]. For Gradient Boosting (GB), the sub-sample parameter is set to 0.95, which trains each base classifier on a fraction of the available data. Sub-sampling is a stochastic behaviour and typically enhances performance.

### 3.8. Combining strokes

Classifiers may struggle to identify each stroke correctly, and it is common to combine strokes when authenticating users. We combine strokes using a moving average over the predicted probabilities, like [14], since it makes the model agnostic by still predicting each stroke separately. Similarly, the feature vector remains the same in contrast to [10]. While the latter approach is also classifier agnostic, we argue that [10] is inadequate for directional features using angles since an average over the left and right direction may result in a behaviour representing neither of the two movements. As such, we use the former method but recognise the merit of the latter in cases where directional features are irrelevant.

## 4. Results and discussion

We first present the classifiers and parameters selected using our approach and with a comparison to the related work. Next, we compare bi- and omnidirectional models in the context of single-stroke authentication before considering the impact of combining strokes. Lastly, we highlight the benefits of combining strokes in the context of the five feature sets and classifiers.

*4.1. Modelling parameters*

We searched through the parameters in Table 3 for the Horizontal (Hs), Vertical (Vs), and Omnidirectional models to better understand which parameters work for most users. For KNN and SVM, the optimal parameter changes depending on the applied feature set and directional modelling approach. For KNN, as seen in Figure 7, most models found three neighbours a suitable parameter when using the TouchAlytics (TA) behaviour. In contrast, the other feature sets change between 1, 3, and 5 neighbours but rarely 7, regardless of direction. While a shared parameter cannon be suggested based on this result, we see a similar difference in the optimal parameter used across the literature, as seen in Table 1, where the optimal *k* range is either 3, 7, 9, or 11, depending on the referenced work.

For SVM and TA, as seen in Figure 8, the Hs models favour $C$=0.1 while the Vs models vary between $C$=0.1 and $C$=1.0. Similar to KNN, different feature sets also prefer different parameter values. However, most models appear to perform decently with a $C$ parameter of 1.0. Similar to the parameter search for SVM, it is challenging to suggest an optimal parameter for all users. Contrary to KNN and SVM, all tree-based classifiers use 100 trees and a sample split of 0.005 for all users, feature sets, and directions. An exception is the GB classifier, with omnidirectional models selecting 200 trees for approximately 5 out of 35 users depending on the feature set. Figure 9 shows an example of the trade-off between performance and time to train an omnidirectional model depending on the parameter complexity selected for the tree-based classifiers. However, we can observe that the GB classifier improves more than the other classifiers when increasing the number of trees but still underperforms compared to the ET classifier.

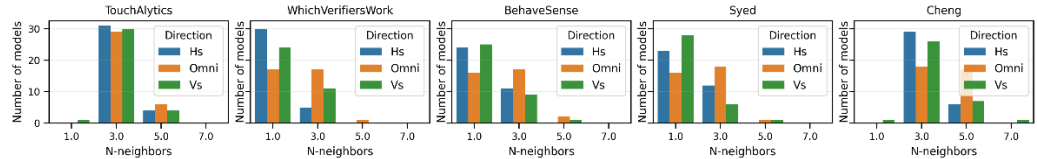

*Figure 7 Frequency of selected parameters for kNN, for each touch direction, using the feature sets from the related work, TouchAlytics (TA), WhichVerifiersWork (WVW), BehaveSense (BS), Syed, and Cheng feature sets,* [6], [9], [12]–[14], *respectively. As shown, the best parameter varies depending on the applied feature set.*

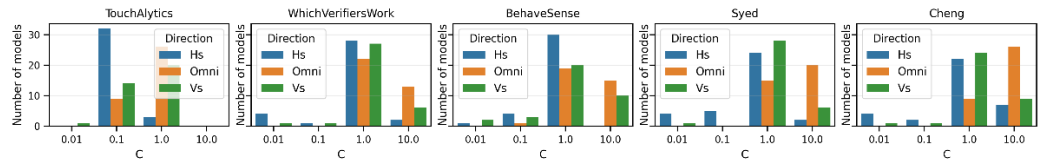

*Figure 8 Frequency of selected parameters for SVM, for each touch direction, using the feature sets from the related work, TA, WVW, BS, Syed, and Cheng feature sets,* [6], [9], [12]–[14], *respectively.*

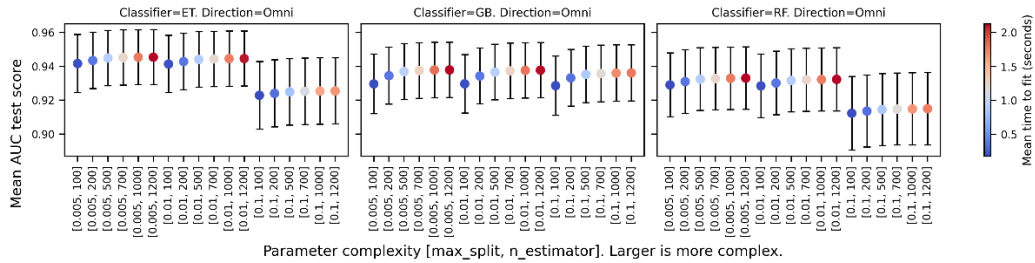

*Figure 9 Mean AUC scores in the context of fitting tree-based models with greater parameter complexity and coloured by the time to train models in seconds. Greater parameter complexity provides limited performance gains while consuming three to four times the time to fit.*

Performance may increase with the number of trees used in a tree-based classifier, but it requires a longer training time. We argue that the performance gain is insignificant, considering it can take up to four times longer to fit the models. Consequently, there is limited benefit in increasing the complexity beyond 100 trees compared to [9], [26], which uses up to 1000. Thus, our parameter selection approach may also benefit the time required to (re)train models.

### 4.2. Bi versus omnidirectional single-stroke comparison

After setting the optimal model parameters, Figure 10 presents the mean AUC score for each classifier, grouped by each feature set when authenticating users using a single stroke with each model. Unsurprisingly, the performance differs amongst the feature sets. BehaveSense (BS) [13] generally ranks top, whilst Syed [14] and WVW [9] often perform poorly.

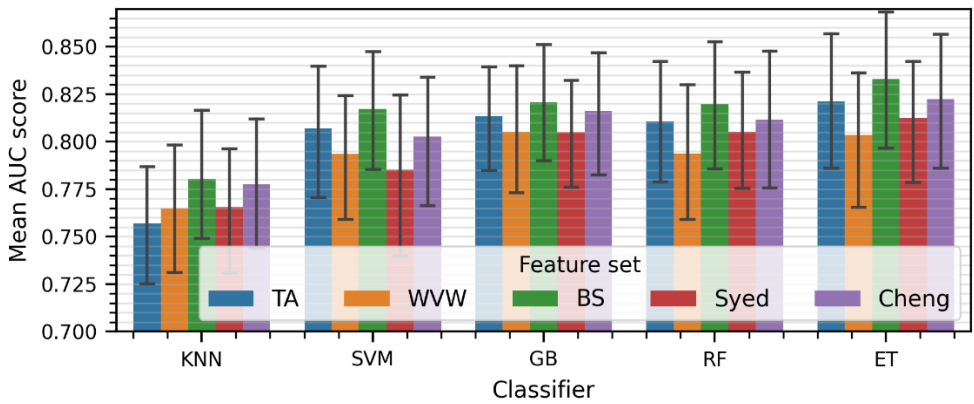

*a) Bidirectional mean AUC scores. The ET classifier generally outperforms others independent of the applied feature set, with the BehaveSense (BS) [13] feature set being superior*

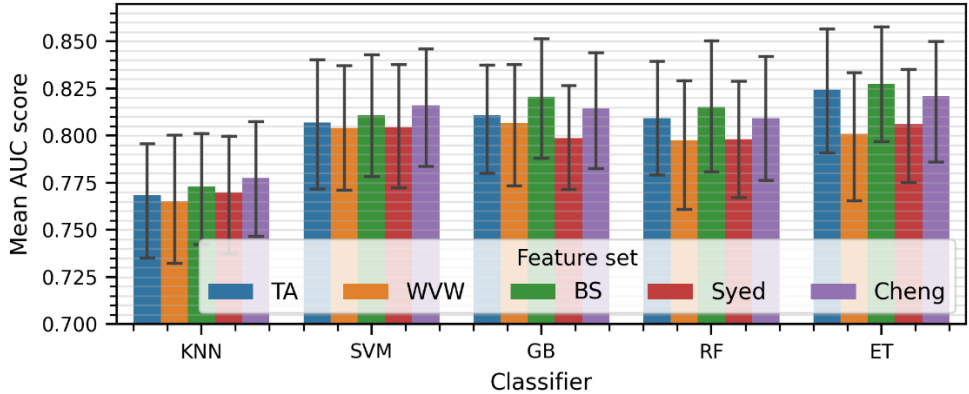

b) *Omnidirectional mean AUC scores. Like the bidirectional approach, the ET classifier generally outperforms others, with the BehaveSense (BS)* [13] *feature set being superior*

*Figure 10 Single-stroke performance. Mean AUC scores across all users for each classifier and feature set while comparing the directional approaches. Error bars indicate the 0.95 Confidence Interval. The Extra Tree classifier outperforms others in combination with the BehaveSense (BS)* [13] *feature set.*

While the bidirectional approach has the highest mean AUC and EER score, the difference from the omnidirectional counterpart is negligible; moreover, the standard deviation for omnidirectional models is slightly lower for both AUC and EER compared to bidirectional models. In [9], they achieved an EER score of 13.8 and 17.2%, Hs and Vs, respectively, but required ten strokes. Similarly, we also notice that some users are more challenging to model than others, as indicated by the wide error bars in Figure 10. Regardless, the goal of this work is not to exclude or identify problematic users but to compare the modelling approach irrespective of these. Table 4 highlights the top five classifiers with the highest mean AUC score. The results for each model are compared against the top-performing model on the first line in Table 4 to detail the answer to research question one further.

*Table 4 Top five single-stroke performances ranked by highest mean AUC score amongst bi and omnidirectional classifiers and feature sets.*

| Classifier | Feature set | Approach | AUC (±STD) | EER (±STD) |
|---|---|---|---|---|
| ET | BS | Bi | 0.833 (±0.103) | 0.239 (±0.098) |
| ET | BS | Omni | 0.827 (±0.098) | 0.247 (±0.094) |
| ET | TA | Omni | 0.824 (±0.096) | 0.247 (±0.087) |
| ET | Cheng | Bi | 0.822 (±0.106) | 0.251 (±0.104) |
| ET | TA | Bi | 0.821 (±0.103) | 0.252 (±0.096) |

### 4.3. Wilcoxon signed-rank test

To measure whether the top-performing model is better than others, we apply the Wilcoxon signed-rank test using the best model as the reference. The test compares the AUC scores between the reference and iteratively selects the other classifiers to evaluate the AUC distributions and whether the best AUC scores differ significantly from any other classifiers.

More specifically, the null hypothesis assumes that the AUC scores predicted by classifier A are from the same distribution as classifier B's. We wish to reject the null hypothesis with a 5% confidence level. If the null hypothesis cannot be rejected, then the reference model is not significantly better than the comparison. Table 5 presents seven classifiers that fail to reject the null hypothesis, suggesting that the traditional Bi approach is not significantly better than the proposed omnidirectional approach.

As such, we can answer the first research question. In the context of single-stroke authentication, there is an insignificant performance difference between the traditional and our proposed omnidirectional methods. While Figure 10 shows that the BS feature set consistently outperforms the others irrespective of the classifier and modelling approach, we note this may not carry over when combining strokes, which the following section covers.

*Table 5 Classifiers that are not significantly different from the single-stroke ET BS Bidirectional AUC distribution*

| Classifier | Feature set | Direction | P-value |
| --- | --- | --- | --- |
| ET | Cheng | Bi | 0.3257 |
| ET | BS | Omni | 0.3098 |
| ET | Cheng | Omni | 0.2013 |
| ET | TA | Omni | 0.1589 |
| ET | TA | Bi | 0.1014 |
| RF | Cheng | Bi | 0.0665 |
| GB | Cheng | Bi | 0.0574 |

### 4.4. Combining strokes

While the best single-stroke classifier was an ET classifier using the BS feature set, we visualise the influence of combining strokes in Figure 11. What stands out is the steady incline in the mean AUC score for the omnidirectional ET classifier using the TA features. Compared to [9], the proposed approach achieves equivalent results using five strokes compared to ten in the referenced work- Table 6 details the top five performing combinations across the bi and omnidirectional methods when combining five strokes. When combining strokes, the best classifier remains an ET classifier, but the feature set changes to TA. Compared to single-stroke authentication results in Table 4, we improved the mean AUC score from 0.833 to 0.890 (+5.7%) and reduced the EER score from 0.239 to 0.179 (-6%). More importantly, the proposed omnidirectional method outperforms the traditional bidirectional approach.

*Table 6 Top five performances, combining five strokes, ranked by highest mean AUC score amongst bi and omnidirectional classifiers and feature sets.*

| Classifier | Feature set | Approach | AUC (±STD) | EER (±STD) |
| --- | --- | --- | --- | --- |
| ET | TA | Omni | 0.890 (±0.099) | 0.179 (±0.112) |
| ET | BS | Bi | 0.886 (±0.106) | 0.181 (±0.112) |
| ET | TA | Bi | 0.886 (±0.109) | 0.182 (±0.117) |
| ET | BS | Omni | 0.881 (±0.096) | 0.190 (±0.104) |
| GB | TA | Bi | 0.881 (±0.093) | 0.190 (±0.103) |

In the context of single-stroke authentication, our approach compares to the traditional one but requires just one model instead of two. Thus, modelling could be faster and easier to manage, deploy, and interpret. At the same time, our approach is superior when combining three strokes or more. We found limited improvements for any methods when combining more than ten strokes. Hence, Figure 11 is limited to combining ten strokes as the curve flattens without changing the rankings of classifiers. Compared to [9], we also combined ten strokes and achieved an average of 0.905 AUC and 0.159 EER score, which is +0.004 EER; however, we have a single model and a more stable standard deviation. We found minimal improvements using more than ten strokes, as seen in Figure 12, which shows omnidirectional performance. The same is true for the bidirectional models combining more than ten strokes. Thus, to answer research questions two and three, we suggest that three to five strokes are enough to provide satisfactory performance. Despite being the earliest feature set, we suggest using the TouchAlytics set since the results show better performance for the ET and amongst many of the classifiers used for bi and omni-directional methods.

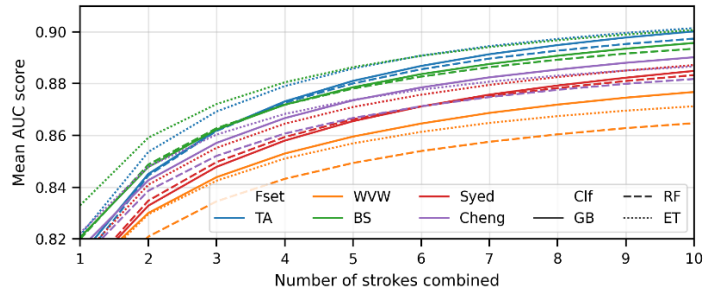

*a) Mean AUC scores for the bidirectional models*

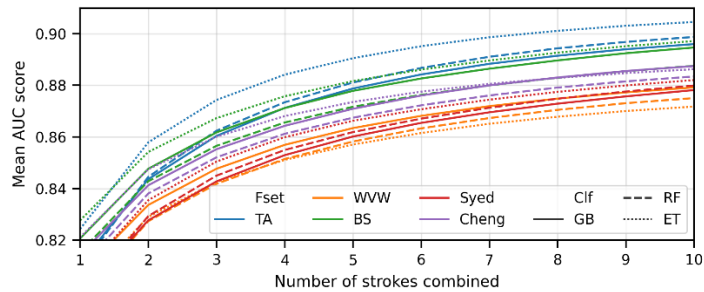

*b) Mean AUC scores for the mean omnidirectional models*

*Figure 11 Overview of the performance impact when combining strokes. The plot is limited to 1 to 10 strokes due to limited gains beyond ten and presents the different classifiers (Clf) – RF, GB, and ET, and the feature sets (Fset) – WVW, Syed, TA, BS, and Cheng. The BS feature set performs well for one stroke but inferior when combining n number of strokes.*

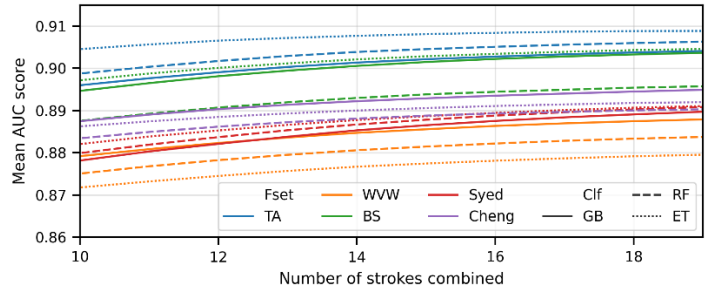

*Figure 12 Mean AUC scores when combining 10 to 20 strokes grouped by classifiers (Clf) – RF, GB, and ET, and the feature sets (Fset) – WVW, Syed, TA, BS, and Cheng, exclusively for the omnidirectional approach. Similar trends appear for the bidirectional method but with slightly lower scores.*

### 4.5. Limitations

#### 4.5.1. Inconsistency of comparable metrics across the literature

AUC is threshold independent and aims to produce models that find the best trade-off between miss-classifying the genuine and non-genuine users. The EER is derived from AUC based on selecting a threshold that separates the two classes while balancing miss-classification equally. However, EER is not the best metric to compare since it depends on the chosen thresholds, which vary between users. Similarly, false acceptance and rejection rates suffer from the same issue. Accuracy is rarely seen in the related work, perhaps since it generalises both true positives and negatives over all data points; thus, a majority class with good performance may skew the results. In our work, we decided to optimise for more significant AUC scores while also providing EER scores to compare with other papers. However, direct comparison with related work is challenging since the metrics are derived using differences such as stroke combining methods, data sub-setting, data cleaning, and user selection.

#### 4.5.2. Feature super and subsets

While this paper focuses on five feature sets from the literature, an evaluation of feature importance can be made to define a superset which combines the best *n* performing features from each related work into a new feature set. Similarly, subsets can be made to eliminate noisy or poorly performing features. However, we took the first steps to compare the feature sets and leave these potential improvements to future work. We highlight that it may be faster to evaluate feature importance using our omni-directional approach since simpler models are faster to train and more straightforward to interpret. For example, new features could be engineered, such as splitting the stroke at the 20 quantiles to better focus on the beginning of touch interactions.

#### 4.5.3. Coordinate specific features

Most feature sets used in touch-based biometrics incorporate at least the start and stop (*x, y*) coordinate pairs as features. However, models relying on coordinate pairs may have a contextual limitation since they can be affected by the screen content. E.g., the placement of a button or other screen content that a user needs to click or when users may avoid covering the screen with their finger while reading. Furthermore, the size of a device may further affect these features despite normalising the coordinates according to the Dots Per Inch, as seen in [6]. This work shows that the BS [13] feature set performed well on single-stroke authentication while suffering when combining strokes. Interestingly, the BS feature set also contains the most coordinate-specific features. It may be better to engineer coordinate-independent features or lean towards the TA [6] feature set.

*4.5.4. Incompatibility between the strokes combining method*

It is challenging to compare results between the state-of-the-art, as the methods to combine strokes differ, e.g., training a model by combining the feature vectors before training [9] or averaging the predicted probabilities [15]. Thus, single-stroke performance should be reported to allow comparisons based purely on model performance, where, under perfect conditions, each stroke could be accurately predicted. However, since models are trained to generalise, it is also essential to examine combining strokes. This work averages the predicted probabilities of classifiers trained on single strokes and a rolling window between 1 (no averaging) to 20 strokes. Thus, the comparison of merging feature vectors before training is left for future work.

*4.5.5. Comparing Omni vs bidirectional paradox*

While the omnidirectional model outperforms the traditional method, a direct comparison may not be fair as the underlying data differs. Specifically, a horizontal model is exposed to 100 strokes, while the omnidirectional must learn the horizontal behaviour collectively from all 200 observations. Hence, our approach may have an advantage in generalisation, which could cause a better performance when combining strokes.

## 5. Conclusion

While the bidirectional models based on an ET classifier work for single-stroke authentication, our approach is comparable and superior when combining three strokes. Interestingly, single-stroke authentication works better using the behaviour captured by the BS feature set, but the TA feature set improves performance when combining strokes. Despite KNN and SVM being commonly used, they are inferior to the tree-based classifiers. We conclude that the omnidirectional approach is preferable when using an ET classifier using the TA feature set and combining at least three strokes. Further, we suggest our hyper-parameter tuning method, providing a lower AUC standard deviation.

## 6. Acknowledgement

The work is supported by the School of Computing - Edinburgh Napier University and The Scottish Informatics & Computer Science Alliance (SICSA).